\documentclass[aps,twocolumn]{revtex4}
\usepackage{graphicx}
\usepackage{epsfig}
\begin{document}
\bibliographystyle{apsrev}
\title{LISA Source Confusion}
\author{Jeff Crowder and Neil J. Cornish}
\affiliation{Department of Physics, Montana State University, Bozeman, MT
59717}

\begin{abstract}
The Laser Interferometer Space Antenna (LISA) will detect thousands of gravitational wave sources. Many of these sources will be overlapping in the sense that their signals will have a non-zero cross-correlation. Such overlaps lead to source confusion, which adversely affects how well we can extract information about the individual sources. Here we study how source confusion impacts parameter estimation for galactic compact binaries, with emphasis on the effects of the number of overlaping sources, the time of observation, the gravitational wave frequencies of the sources, and the degree of the signal correlations. Our main findings are that the parameter resolution decays exponentially with the number of overlapping sources, and super-exponentially with the degree of cross-correlation. We also find that an extended mission lifetime is key to disentangling the source confusion as the parameter resolution for overlapping sources improves much faster than the usual square root of the observation time.
\end{abstract}

\maketitle

\section{Introduction}

It is anticipated that the Laser Interferometer Space Antenna (LISA)~\cite{PrePhaseA} will start collecting data in the next decade. LISA will operate in the source-rich low frequency portion of the gravitational wave spectrum, where the primary sources are expected to be compact galactic binaries, supermassive black hole binaries, and extreme mass ratio inspirals (such as a white dwarf falling into a supermassive black hole). A natural question to ask is how well the source parameters, such as sky location, masses and distances, can be recovered from the LISA data stream. Estimates of the parameter resolution have been given for compact galactic binaries\cite{Cutler,Hellings,Seto1}, supermassive black hole binaries\cite{Cutler,Hellings,Hughes,Seto2,Vecchio}, and extreme mass ratio inspirals\cite{Barack}. These studies focused on the problem of identifying one source at a time, and did not address the problem of source confusion. The large signal from the galactic population of close white dwarf binaries was not ignored, but it was treated as an additional source of stationary, Gaussian noise to be added to the instrument noise. We will see that this is not a good approximation.

Here we study how parameter estimation is affected by the presence of multiple overlapping signals in the LISA data stream. Importantly, we find that the behavior is very different from what one would predict by treating the additional sources as stationary, Gaussian noise. The problem of source confusion will be most pronounced below $\sim 2$ mHz, where it is expected that many tens to tens of thousands of galactic binaries will have signals that overlap in each frequency bin~\cite{H_B_W}. Thus, we will focus our attention on low frequency galactic binaries that are close to one another in frequency.

The paper is organized as follows. We begin with a summary of our findings in Section~\ref{sum}. This is followed by a review of parameter estimation for an isolated galactic binary in Section~\ref{rev}. Parameter estimation with multiple sources is described in Section~\ref{multi}, and our plan for exploring the parameter space is described in Section~\ref{param}. The main results are described in Section~\ref{main}. A brief information theory perspective is presented in Section~\ref{info}, and concluding remarks are made in Section~\ref{concl}.

\section{Summary of Results}\label{sum}
We find that the parameter estimation uncertainties grow exponentially with the number, $N$, of overlapping sources. This should be contrasted to the $\sqrt{N}$ increase one would predict if the other sources were treated as stationary, Gaussian noise. The degredation in resolution was found to be nearly uniform across the seven parameters that LISA will measure for a galactic binary system. As one might expect, the parameter uncertainties are a strong function of the signal cross-correlations. We find that as the two signals become more correlated, the parameter estimation uncertainties grow at a rate that is faster than exponential. Importantly, we find that the parameter uncertainties decrease rapidly as the observation time $T$ is increased - far more rapidly than the usual $1/\sqrt{T}$ improvement one expects when competing with stationary, Gaussian noise.

\section{Review of Parameter Estimation}\label{rev}

Typical galactic binaries can be treated as circular and monochromatic. They are described by seven parameters: sky location ($\theta, \phi$); gravitational wave frequency $f$; amplitude $A$;  inclination and polarization angles ($\iota, \psi$); and the initial orbital phase $\gamma$. The response of the LISA instrument to a gravitational wave source is encoded in the (Micheleson-like) $X(t)$ and $Y(t)$ time-delay interferometry variables~\cite{tdi}. We employ the rigid adiabatic approximation~\cite{rcp} to describe these variables, and work with the orthogonal combinations~\cite{Cutler}

\begin{equation}
S_{I}(t) = X(t), \quad S_{II}(t) = \frac{1}{\sqrt{3}}\left(X(t)+2
Y(t)\right)\, .
\end{equation}

Each signal is a function of the parameters $\vec{\lambda} \rightarrow (\theta,\phi, f, A, \iota, \psi, \gamma)$ that describe the source. Denoting the noise spectral power by $S_n(f)$, we adopt the usual noise-weighted inner product

\begin{equation}
\langle a \vert b \rangle = 2 \int_0^\infty
\frac{ a^*(f) b(f) + a(f) b^*(f)}{S_n(f)} \, df .
\end{equation}

Decomposing the output channels into signal $h$ and noise $n$, the signal to noise ratio is given by

\begin{equation}
{\rm SNR}^2 = \sum_{\alpha = I,II} \langle h_\alpha \vert h_\alpha
\rangle .
\end{equation}

For large SNR the parameter estimation uncertainties $\Delta \lambda^i$ will have the Gaussian probability distribution~\cite{Cutler_and_Flannigan}

\begin{equation}
p(\Delta \lambda^i) = \sqrt{\frac{{\rm det} (\Gamma)}{2\pi}}
\exp\left(-\frac{1}{2}\Gamma_{ij}
\Delta \lambda^i \Delta \lambda^j\right) ,
\end{equation}
where the Fisher information matrix $\Gamma$ is defined by

\begin{equation}
\Gamma_{ij} = \sum_{\alpha = I,II} \langle h_{\alpha ,i} \vert h_{\alpha,
j} \rangle \, .
\end{equation}

Here $h_{, i} = \partial h / \partial \lambda^i$. For large SNR the variance-covariance matrix is given by

\begin{equation}
C^{ij} = \left( \Gamma^{-1} \right)^{ij}
\end{equation}
and the uncertainties in the parameters are given by $\Delta \lambda^i = (C^{ii})^{1/2}$. The volume of the $n-\sigma$ uncertainty ellipsoid in the $d$ dimensional parameter space is

\begin{equation}\label{vol}
V_d = \frac{\pi^{d/2}n^d}{\Gamma(d/2+1)}\sqrt{{\rm det}\, C_{ij}} \, .
\end{equation}

\begin{figure}
\includegraphics[angle=270,width=0.48\textwidth]{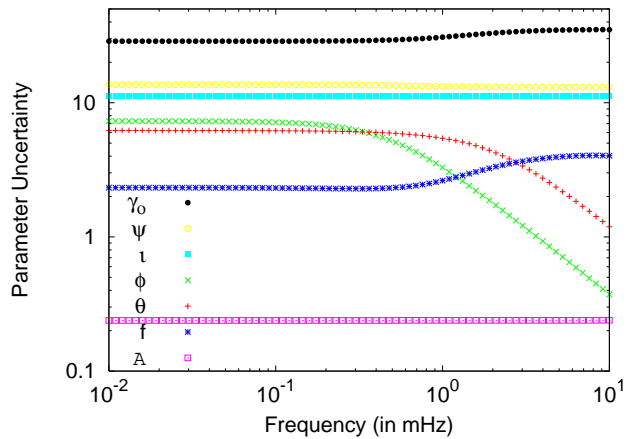}
\caption{\label{Single_source_frequency_plot}Median parameter uncertainties versus frequency for isolated monochromatic binary sources with one year of observation and fixed signal-to-noise ratios of ${\rm SNR}=10$. The parameters are $f, \ln A, \theta, \phi, \iota, \psi, \gamma$.}
\end{figure}

\begin{figure}
\includegraphics[angle=270,width=0.48\textwidth]{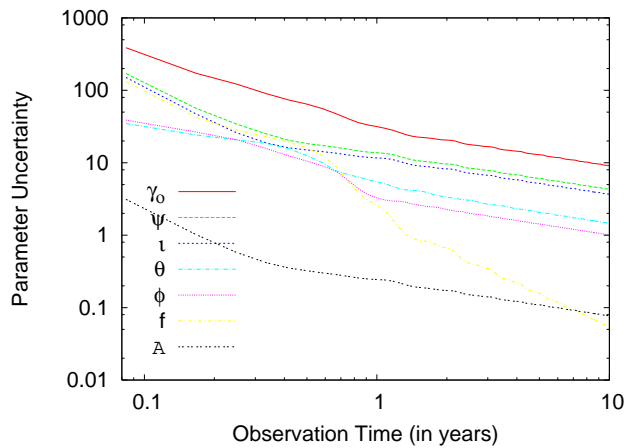}
\caption{\label{Single_source_time_plot}Parameter uncertainties versus time for isolated monochromatic binary sources with $f= 1$ mHz and fixed signal-to-noise ratios of ${\rm SNR}=10$. The parameters are $\ln f, \ln A, \theta, \phi, \iota, \psi, \gamma$.}
\end{figure}

Applying the above formalism to an isolated galactic binary yields the parameter uncertainties shown in Figures~\ref{Single_source_frequency_plot} and~\ref{Single_source_time_plot}. The plots were generated by taking the median values for $10^5$ different sources, each normalized to a signal-to-noise ratio of ${\rm SNR} = 10$. The uncertainty in the frequency has been increased by a factor of $10^6$ for plotting purposes.

As can be seen in Figure~\ref{Single_source_frequency_plot}, the uncertainties in five of the seven parameters are fairly constant across the LISA band. The two exceptions are the sky location variables $(\theta,\phi)$, which show a marked decrease in their uncertainties above 1 mHz. This behavior can be traced to the time varying Doppler shift, which is one of the two ways LISA is able to locate sources on the sky. The Doppler shift increases linearly with $f$, which translates into a $1/f$ decrease in the positional uncertainties above 1 mHz. Below this frequency the the angular resolution comes mainly from the time varying antenna sweep, which is weakly dependent on $f$ below the transfer frequency $f_* \simeq 10$ mHz.

\begin{figure}
\begin{tabular}{c}
      \resizebox{0.48\textwidth}{!}{\includegraphics[angle=270, width=0.48\textwidth]{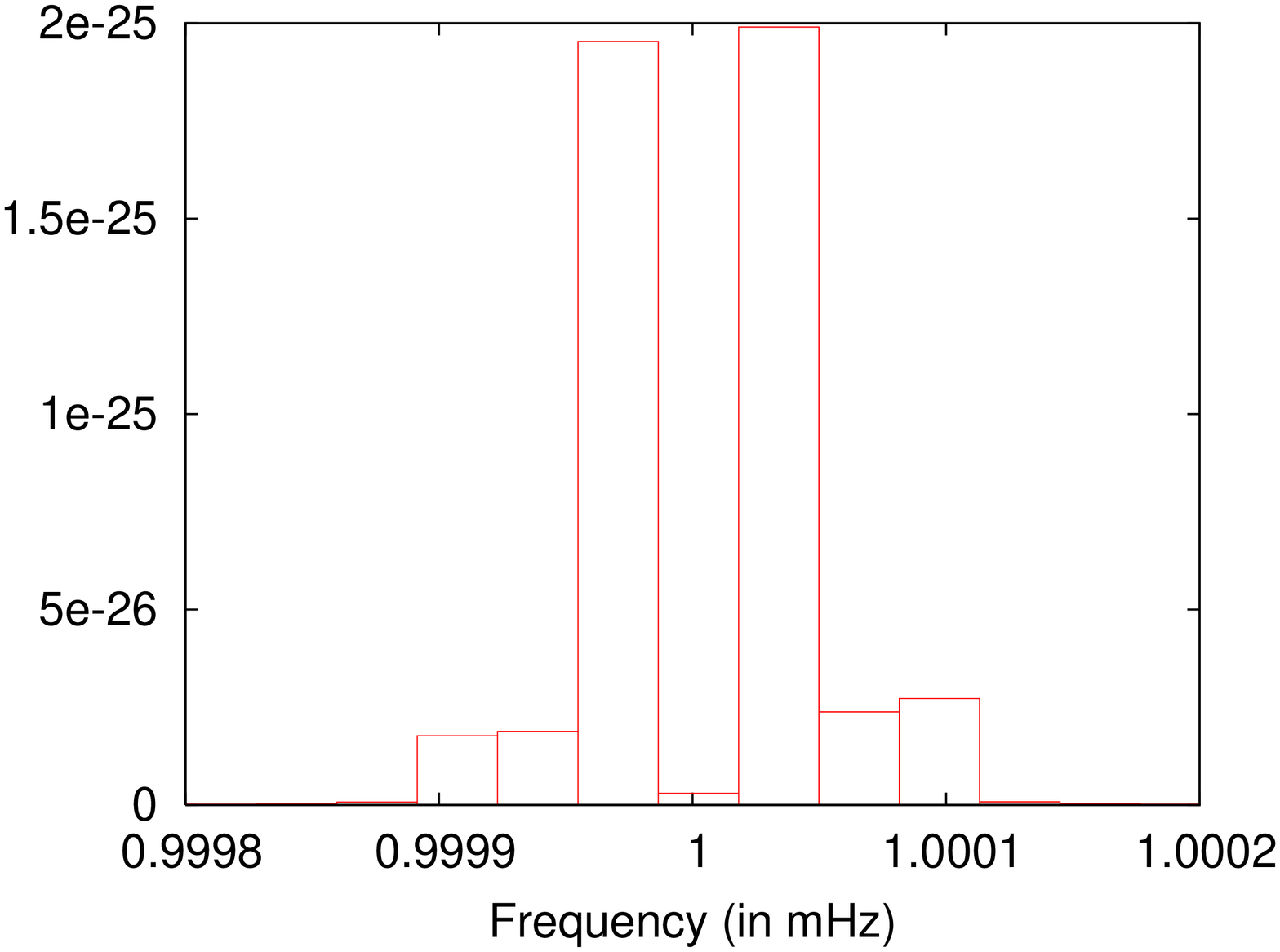}} \\
      \resizebox{0.48\textwidth}{!}{\includegraphics[angle=270, width=0.48\textwidth]{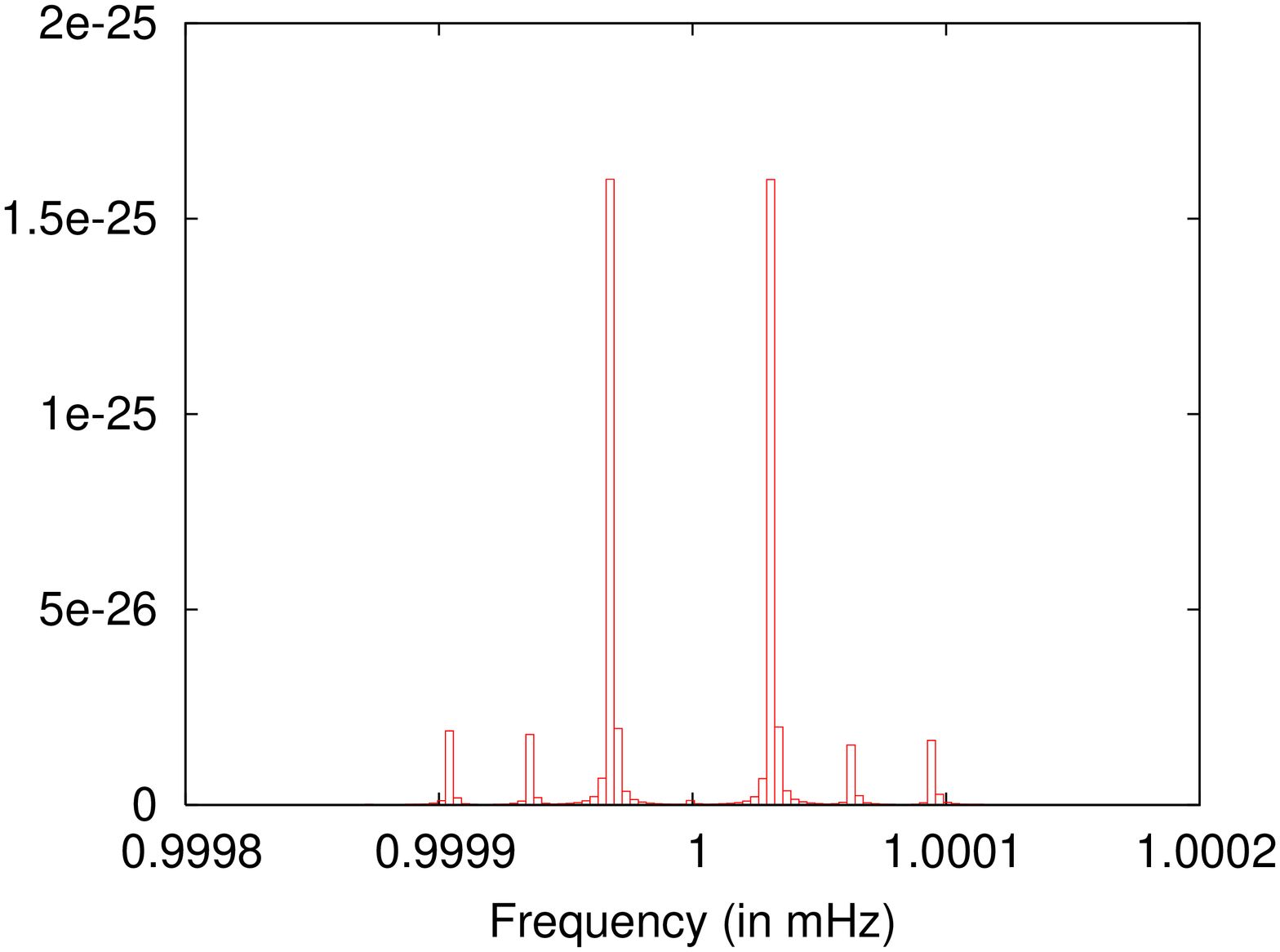}} \\
\end{tabular}
\caption{\label{fftw}Fourier transforms of a typical signal for observation times of one and ten years.}
\end{figure}

As can be seen in Figure~\ref{Single_source_time_plot}, the uncertainties in six of the seven parameters decrease as $1/\sqrt{T_{obs}}$ for observation times longer than a year, in accordance with the standard $\sqrt{T_{obs}}$ increase in the SNR. The one exception is the frequency, which benefits from an additional $1/T_{obs}$ shrinkage in the width of the frequency bins, leading to an overall $1 / T_{obs}^{3/2}$ decay in the frequency uncertainty. The shrinkage in the size of the frequency bins is illustrated in Figure~\ref{fftw}, where the discrete Fourier transform of a typical signal is shown after one year and after ten years of observation. Later we will see how this improved resolution of the sidebands helps reduce source confusion when multiple overlapping sources are present. For observation times less than one year the positional uncertainties also decay as $1 / T_{obs}^{3/2}$. The additional factor of $1/T_{obs}$ can be traced to the two mechanisms by which LISA can locate a source:  antenna sweep and Doppler shift. The angular resolution due to antenna sweep depends on the angle through which the antenna has been swept, which increases as $T_{obs}$ for $T_{obs} < {\rm year}$. The angular resolution due to the variable Doppler shift can be thought of as a form of long baseline interferometery. The resolution improves with the length of the synthesized baseline, which grows as $T_{obs}$ for $T_{obs} < {\rm year}$.

In what follows, we will typically quote our results in terms of uncertainty ratios. These ratios will compare the parameter resolution when one or more overlapping sources are present to the parameter resolution that would be possible if each source were isolated. To arrive at the absolute uncertainties one needs to multiply by the isolated source uncertainties quoted above.

\section{Multiple Sources}\label{multi}

The Fisher information matrix approach to estimating parameter uncertainties is easily generalized to multiple sources. For $N$ circular, monochromatic binary systems, the parameter space is $7N$ dimensional, and we have the parameter vector

\begin{equation}
\vec{\Lambda} =\vec{\lambda}^1 + \vec{\lambda}^2 + \dots +\vec{\lambda}^N
\, ,
\end{equation}

We will use the notation $\Lambda^A$ to refer to all 7 parameters of the $A^{\rm th}$ binary.  The total signal $S=S_I$ or $S=S_{II}$ will be the sum of the signals for the individual binary systems,

\begin{equation}
S = S^1 + S^2 + S^3 \ldots + S^N
\end{equation}

Here $S^A$ is the signal due to the $A^{\rm th}$ binary. Since $\partial S^A / \partial \Lambda_B = 0$ for $A\neq B$, the full Fisher information matrix has a simple structure. If we define

\begin{equation}
\Gamma_{AB} = \sum_{\alpha = I,II}
\langle \frac{\partial S^A_\alpha}{\partial \Lambda_A} \vert
\frac{\partial S^B_\alpha}{\partial \Lambda_B} \rangle \, ,
\end{equation}
the diagonal blocks $\Gamma_{AA}$ are the usual $7 \times 7$ Fisher information matrix for the $A^{\rm th}$ source, while the off-diagonal blocks $\Gamma_{AB}$ describe how the parameter estimation for source $A$ is influenced by the presence of source $B$ and vice-versa. If all the sources are un-correlated (for example, they might be widely spaced in frequency), then the off-diagonal blocks will be zero, and the full Fisher information matrix will be block diagonal. Upon inverting to get the variance-covariance matrix the block diagonal structure will be preserved, with each $7 \times 7$ block equal to the inverse of the corresponding single source Fisher information matrix. However, if the sources are overlapping, the off-diagonal blocks will be non-zero, which will affect the values of the diagonal elements in the full variance-covariance matrix. The number of off-diagonal elements grows quadratically with the number of sources, while the number of diagonal elements grows linearly with the number of sources. Thus, we anticipate that the ability to resolve a particular source's parameters will degrade rapidly as the number of correlated sources increases.

We also anticipate that the parameter uncertainties will depend strongly on the degree of correlation between sources. The correlation between two signals $S^A$ and $S^B$ is defined:

\begin{equation}\label{corr}
\kappa_{AB} = \frac{ \langle S^A \vert S^B \rangle}{\langle S^A \vert S^A
\rangle^{1/2}
\langle S^B \vert S^B \rangle^{1/2} }
\end{equation}

	In analyzing the results, two similar approaches will be used to obtain a quantitative measure of the increase in the parameter estimation uncertainties. First is a global comparison of the uncertainties. The uncertainties form an ellipsoid in the parameter space whose volume is given by the determinant of the variance-covariance matrix. One measure of the uncertainty increase due to the correlation of the binary systems is given by a ratio of the geometric mean of the uncertainties (GMUR):

\begin{equation}
{\rm GMUR} \equiv \left( \frac{{\rm det}\, \Gamma}{\prod_A {\rm det}\, 
\Gamma_{AA}} \right)^{1/7N}
\end{equation}
The GMUR describes the mean increase in the parameter uncertainties due to source confusion.

	A second measure of the uncertainty increase is a parameter by parameter comparison, looking at the ratio of the parameter uncertainty (PUR) between the uncertainty of a particular parameter calculated in the $7N$-dimensional variance-covariance matrix, $\Delta\Lambda^A_x$, and the uncertainty of the same parameter calculated in the isolated binary's variance-covariance matrix, $\Delta\lambda^A_x$.

\begin{equation}
{\rm PUR} \equiv \frac{\Delta\Lambda^i_x} {\Delta\lambda^i_x} \,.
\end{equation}

This definition for the PUR is independent of the amplitudes of the sources, as can be seen in the simple case of two sources, each described by one $\alpha$.  The PUR for $\alpha_1$ is given by:

\begin{equation}
{\rm PUR}_{\alpha_1} = \frac{\Delta\Lambda_{\alpha_1}}
{\Delta\lambda_{\alpha_1}} =
\frac{1}{\sqrt{1-\Sigma_{\alpha_1\alpha_2}^2}}\, ,
\end{equation}
where

\begin{equation}
\Sigma_{\alpha_1\alpha_2} \equiv \frac{
\Gamma_{\alpha_1\alpha_2} } { \sqrt{\Gamma_{\alpha_1\alpha_1}
\, \Gamma_{\alpha_2\alpha_2}}} \, .
\end{equation}

The independence of the degree of confusion on the source amplitudes would seem to imply that an arbitarily weak source can affect parameter estimation to the same degree as an arbitarily strong source. Clearly, this makes no sense in the limit that the amplitude of the weak source goes to zero. The resolution to this apparent paradox is that the Fisher information matrix approach is only meaningful for sources with ${\rm SNR} > 1$, so arbitarily weak sources are not permitted in the analysis.

\section{Source Parameter Selection}\label{param}

The parameter space for $N$ slowly evolving binaries is $7N$ dimensional. This large dimensionality makes it difficult to carry out an exhaustive exploration of all the circumstances that can affect parameter estimation. We focused on sources that were close in frequency, and chose the other source parameters randomly. The frequency chosen for the first binary fixes the base frequency $f_{\rm base}$ of a data run. The frequencies of the remaining $N-1$ binaries were assigned frequencies of $f_{\rm base}+(i+x)f_m$, where $f_m=1/{\rm year}$ is the modulation frequency, $i$ is an integer, and $x$ is a random number between $0$ and $1$. Thus the remaining binaries are between $i$ and $i+1$ modulation frequency units from the base frequency. Sky locations were chosen by two methods, the first being a random draw on $\cos(\theta)\in [-1,1]$ and $\phi \in [0,2\pi)$ with a fixed distance to the binary of $1$ kiloparsec, the second being a random draw from a galactic distribution~\cite{H_B_W} of $\theta$, $\phi$, and binary distance. The values of the polarization $\psi$ and orbital phase $\gamma_o$ were randomly drawn between $0$ and $\pi$ and $0$ and $2\pi$ respectively. The inclination $\iota$ was taken from a random draw on $\cos(\iota)\in [-1,1]$, but with values outside of the range $\iota \in [1^\circ, 179^\circ]$ rejected to avoid the degeneracy in the gravitational wave produced by a circular binary viewed along its axis. Masses were taken to equal $0.5M_\odot$, as we wished to model white dwarf binaries. 

\section{Results}\label{main}

\subsection{Varying time of observation}\label{time}

\begin{figure}
\includegraphics[angle=270,width=0.48\textwidth]{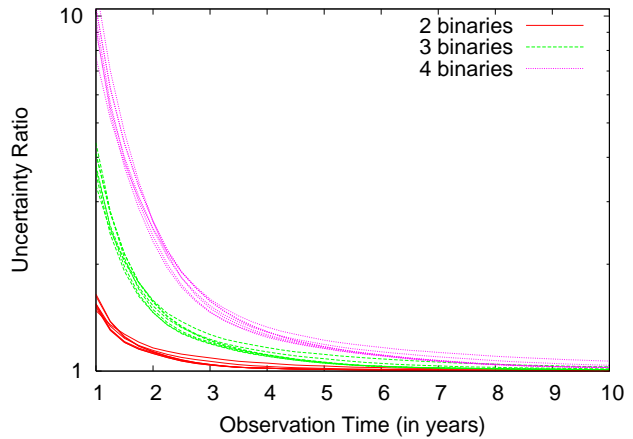}
\caption{\label{parameter_uncertainty_vs_observation_time}Median PUR versus time of observation for two, three, and four binaries with base frequencies of $1$ mHz from an all sky draw.}
\end{figure}

\begin{figure}
\includegraphics[angle=270,width=0.48\textwidth]{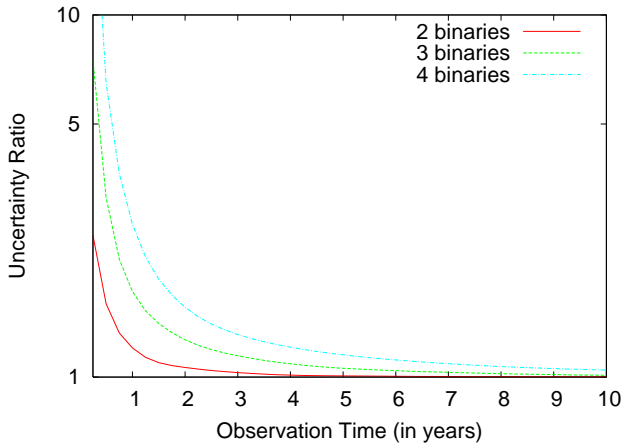}
\caption{\label{volume_uncertainty_vs_observation_time}Median GMUR versus time of observation for two, three, and four binaries with base frequencies of $1$ mHz from an all sky draw.}
\end{figure}

	The current lifespan of LISA is estimated to be five years. Our data show that the longer LISA observes the binaries systems, the less the effect of confusion between sources. Figure~\ref{parameter_uncertainty_vs_observation_time} shows the median parameter uncertainty ratios (PURs) for two, three, and four binary systems with base frequencies of $1$ mHz as a function of the time of observation. Figure~\ref{volume_uncertainty_vs_observation_time} shows the median geometric mean uncertainty ratio (GMUR) for the same data. Each set of PUR data in Figure~\ref{parameter_uncertainty_vs_observation_time} contains the uncertainty ratios for all seven parameters, and as can be seen in the figures, the uncertainty ratios for each of the seven parameters are closely related. This relationship holds true for other frequencies in the LISA band. It is this uniformity in the increase in parameter uncertainties that motivates our use of the GMUR as an estimate for the individual PURs.  Thus, subsequent results will be given in terms of the GMUR.

\begin{figure}
\includegraphics[angle=270,width=0.48\textwidth]{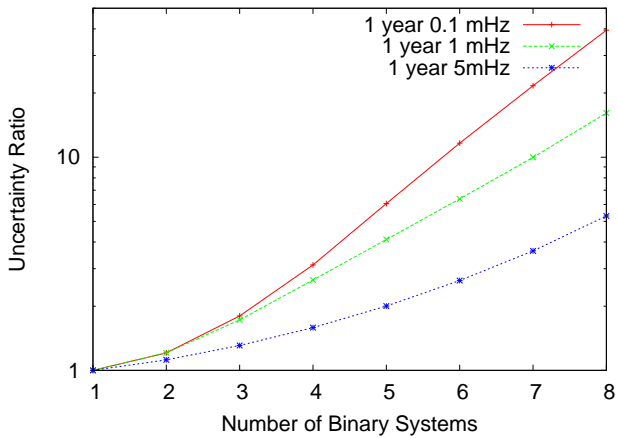}
\caption{\label{GMUR_vs_number_of_binaries_one_year_multiple_freqs}Median GMUR plotted against the number of binary systems for one year of observation at base frequencies of $0.1$, $1$, and $5$ mHz.}
\end{figure}

\begin{figure}
\includegraphics[angle=270,width=0.48\textwidth]{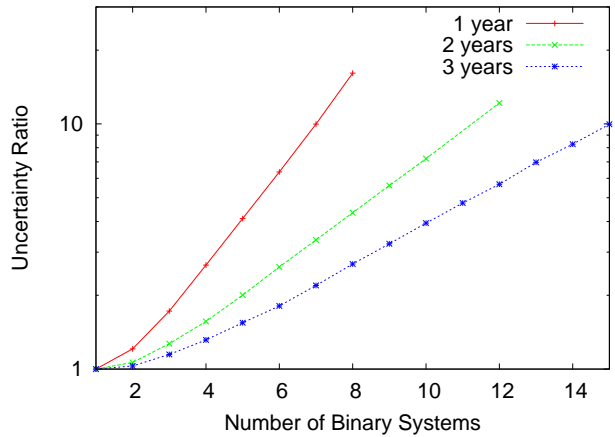}
\caption{\label{GMUR_vs_number_of_binaries}Median GMUR plotted against the number of binary systems for one, two, and three years of observation at a base frequency of $1$ mHz.}
\end{figure}

	Figure~\ref{GMUR_vs_number_of_binaries_one_year_multiple_freqs} shows how the median GMUR increases with the number of binaries for one year of observation, at base frequencies of $0.1$, $1$, and $5$ mHz. Figure~\ref{GMUR_vs_number_of_binaries} shows how extending the time of observation affects the GMUR. In both instances, the parameter uncertainties increase roughly exponentially with the number of overlapping sources.

\begin{figure}
\includegraphics[angle=270,width=0.48\textwidth]{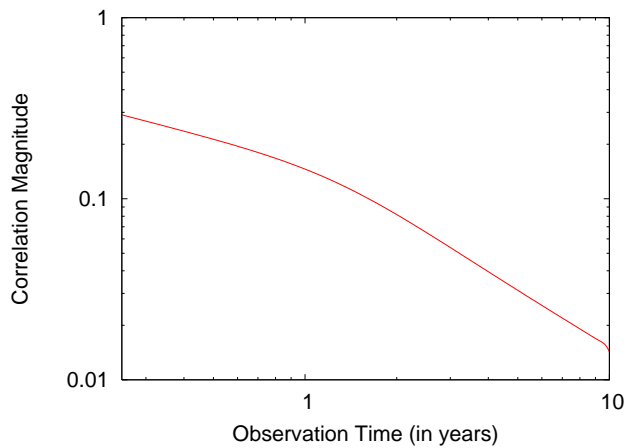}
\caption{\label{correlation_vs_observation_time}Median Correlation magnitude versus time of observation for two binaries with a base frequency of $1$ mHz from an all sky draw.}
\end{figure}

	Figure~\ref{correlation_vs_observation_time} shows the median of the magnitude of the correlation between the signals of two binary systems with a base frequency of 1 mHz taken from an all sky draw. As expected, the correlation magnitude falls off as  $1 / T_{obs}$ for observation times greater than a year (the numerator in (\ref{corr}) oscillates, while the denominator grows as $T_{obs}$). The $1 / T_{obs}$ fall-off in the signal correlation should be contrasted with the much faster fall-off in the GMUR seen in Figure~\ref{GMUR_vs_number_of_binaries}. The reason for this difference in fall-off will become clear in Section~\ref{subsection_c}.

\subsection{Uncertainty ratios for two binary systems as a function of frequency difference}

\begin{figure}
\includegraphics[angle=270,width=0.48\textwidth]{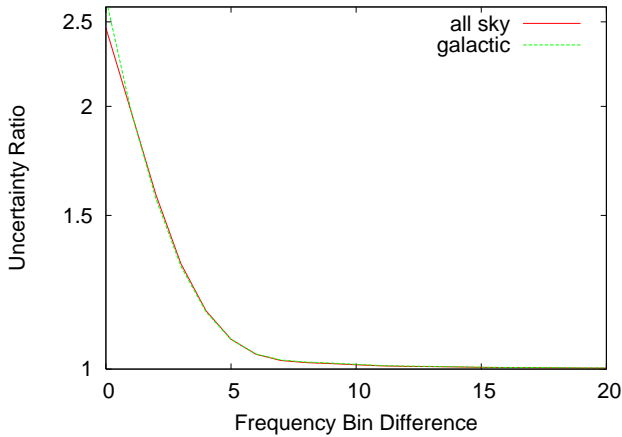}
\caption{\label{volume_ratio_vs_bin_difference}Median GMUR plotted against modulation frequency bin difference of two binaries from an all sky and a galactic draw with a base frequency of $1$ mHz and one year of observation}
\end{figure}

Figure~\ref{volume_ratio_vs_bin_difference} shows how the median parameter uncertainties depend on the frequency separation when two binary systems are present. The base frequency, $f_{base}$, for the first binary was held fixed at 1 mHz, while the frequency of the second binary was randomly chosen between $f_{base}$ and $f_{base} + i f_m$ for $i$ between $0$ and $20$. Figure~\ref{volume_ratio_vs_bin_difference} shows the plots for GMUR versus $i$ for all sky and galactic draws for one year of observation.

The uncertainty ratio drops rapidly as the frequency separation increases, approaching unity by the time the binaries are $\sim 5 f_m$ apart. The frequency difference of $\sim 5 f_m$ corresponds to the typical half-bandwidth of a source at 1 mHz. In other words, 1 mHz sources seperated by $> 5 f_m$ have almost no overlap, and thus there is little source confusion for $i>5$.

\subsection{\label{subsection_c}Uncertainty ratios as a function of signal correlation}

\begin{figure}
\includegraphics[angle=270,width=0.48\textwidth]{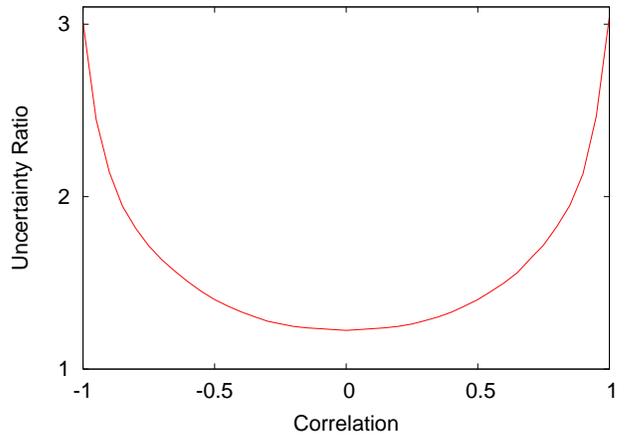}
\caption{\label{volume_uncertainty_vs_correlation_gx} The median GMUR plotted against signal correlation of two binaries from a galactic sky draw with a base frequency of $1$ mHz and one year of observation}
\end{figure}

\begin{figure}
\includegraphics[angle=270,width=0.48\textwidth]{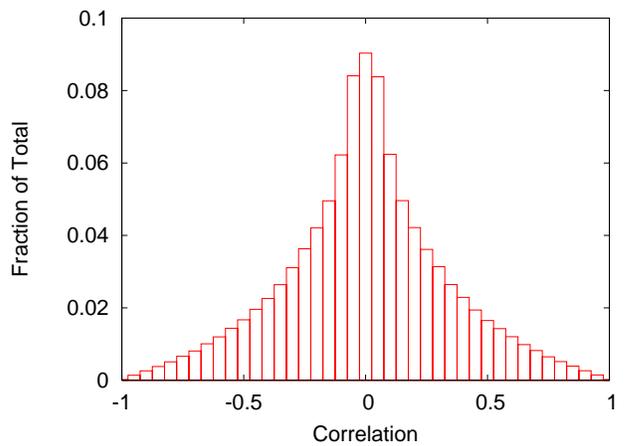}
\caption{\label{bin_count_vs_correlation} A histogram the correlation of two binaries from a galactic draw with a base frequency of $1$ mHz and one year of observation}
\end{figure}

\begin{figure}
\includegraphics[angle=270,width=0.48\textwidth]{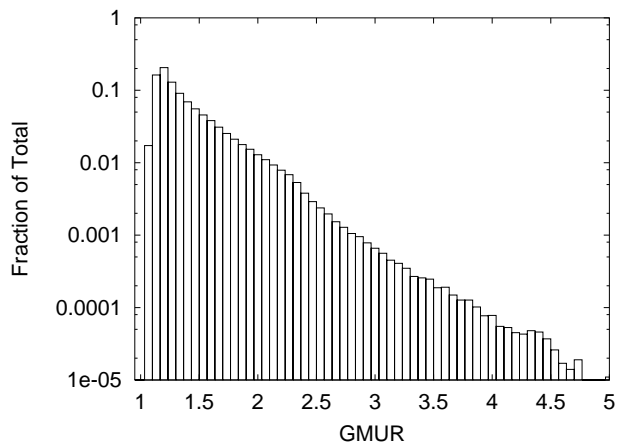}
\caption{\label{bin_count_vs_GMUR} A histogram the GMUR of two binaries from a galactic draw with a base frequency of $1$ mHz and one year of observation}
\end{figure}

	Figure~\ref{volume_uncertainty_vs_correlation_gx} plots the median GMUR versus the signal correlation for a galactic draw when two binaries are present within $\delta f = f_m$ of each other. The plot shows that the GMURs increase faster than exponentially as the signal correlation increases. While a link between signal correlation and uncertainty ratios is to be expected, the super-exponential nature of the relationship came as a suprise. Figures~\ref{bin_count_vs_correlation} and \ref{bin_count_vs_GMUR} show the spreads for correlation and GMUR in the same data run.  As can be seen in the log-linear scale of Figure~\ref{bin_count_vs_GMUR}, the GMUR falls off exponentially from its most frequent value of 1.2. 

While Figure~\ref{bin_count_vs_correlation} shows the results for two correlating binaries, its results can be extended to multiple correlating binaries.  For example, in the figure one can see that $87\%$ of the pairs of binaries have correlation magnitudes $| \kappa | < 0.5$.  Thus with $N$ binaries the probability that there will be a pair of binaries with $| \kappa | > 0.5$ grows as $1-0.87^{(N (N-1) / 2)}$.  As binaries overlapping in frequency are expected to number from tens to tens of thousands, the probability of having high correlation is very close to 1.

The rapid increase in the parameter uncertainties with signal correlation explains the results seen in Section~\ref{time}. The $1 / T_{obs}$ decrease in the signal correlation translates into a much faster decrease in the GMUR as a function of observation time.

\subsection{Uncertainty ratios as a function of base frequency}

\begin{figure}
\includegraphics[angle=270,width=0.48\textwidth]{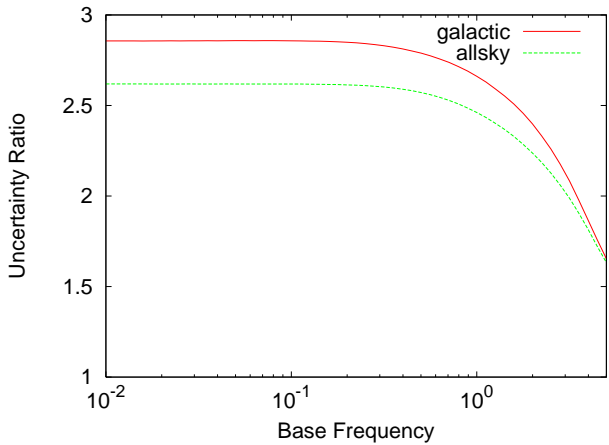}
\caption{\label{volume_uncertainty_vs_freq}Median GMUR plotted against base frequency of two binaries from all sky and galactic draws with one year of observation }
\end{figure}

Figure~\ref{volume_uncertainty_vs_freq} shows how the parameter uncertainty ratios depend on base frequency. The plots are for all sky and galactic draws with two binaries for one year of observation. The base frequency, $f_{base}$, was varied between $0.01$ and $10$ mHz, while the frequency of the second binary was randomly chosen between $f_{base}$ and $f_{base}$ + $f_m$. The GMUR was found to be fairly constant below 1 mHz, followed by a rapid decrease at frequencies above 1 mHz. This behavior can be traced to the different Doppler shifts experienced by each source. The motion of LISA relative to a source located at $(\theta,\phi)$ imparts a Doppler shift equal to

\begin{equation}
\delta f_D \simeq \pi \sin \theta \sin(2\pi f_m t -\phi)\, \left(\frac{f}{{\rm mHz}}\right) f_m \, .
\end{equation}

\begin{figure}
\includegraphics[angle=270,width=0.48\textwidth]{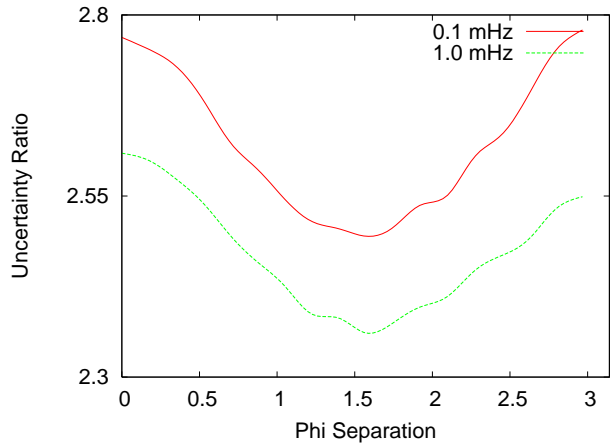}
\caption{\label{gmur_v_psp}Median GMUR plotted against azimuthal sky separation for two pairs of binaries from a galactic draw with one year of observation with base frequencies of $0.1$ and $1$ mHz.}
\end{figure}

The magnitude of $\delta f_D$ becomes comparable to the frequency resolution $\delta f = 1/T_{obs}=1/{\rm year}$ for $f\sim 0.3$ mHz. Sources that are well seperated in ecliptic azimuth $\phi$ will experience Doppler shifts that differ in sign, as LISA will be moving toward one source and away from the other. Thus, we expect that the degree of source confusion will depend on the azimuthal separation of the two sources. Our expectations are confirmed in Figure~\ref{gmur_v_psp}, where the dependence of the GMUR on azimuthal separation is plotted for base frequencies of 0.1 and 1 mHz. Unfortunately, most galactic sources are within $\sim 20^\circ$ degrees of each other in ecliptic azimuth, which helps explain the larger GMURs for the galactic distribution as compared to the all-sky distribution.

\section{Information Theory Perspective}\label{info}

There have been several attempts to understand LISA source confusion in the framework of information theory~\cite{Cornish,Phinney,Hellings_2}. According to Shannon~\cite{shan}, the maximum amount of information that can be transmitted over a noisy channel with bandwidth $B$ in a time $T_{{\rm obs}}$ is

\begin{equation}
I_S = B\, T_{{\rm obs}} \log_2\left( 1 + \frac{P_h}{P_n} \right).
\end{equation}

Here $P_h$ and $P_n$ are, respectively, the signal and noise power across the bandwidth. (Note: The SNR ratio is related to these quantites by ${\rm SNR}^2 \simeq B\, T_{{\rm obs}} P_h /P_n$). As an example, a typical bright galactic source at 1 mHz with $P_h/P_n \sim 10$ can transmit at most $I_S \simeq 70$ bits of information in one year via the two independent LISA data channels $S_I$ and $S_{II}$. In practice the actual information content will be somewhat less due to sub-optimal encoding. It takes $\log_2(x/\Delta x)$ bits of information to store a number $x$ to accuracy $\Delta x$, thus the total amount of information required to describe a monochromatic binary to a precision $\Delta \vec\lambda$ is

\begin{eqnarray}
 && I_\Delta = \log_2\left(\frac{f}{\Delta f}\right) +
\log_2\left(\frac{A}{\Delta A}\right) +
\log_2\left(\frac{\pi}{\Delta \theta}\right)  \nonumber \\
&& \quad \quad + \log_2\left(\frac{2\pi}{\Delta \phi}\right)
+\log_2\left(\frac{\pi}{\Delta \psi}\right) \nonumber \\
&& \quad \quad + \log_2\left(\frac{\pi}{\Delta \iota}\right) +
\log_2\left(\frac{2\pi}{\Delta \gamma}\right) 
\end{eqnarray}

Using the results in Figure~\ref{Single_source_frequency_plot} for a bright galactic source at 1 mHz with $P_h/P_n \sim 10$, we find that $I_\Delta \simeq 45$, which indicates that the encoding, while sub-optimal, is quiet respectable.

One might hope that information theory could be used to predict some of the other results described in Section~\ref{main}. Unfortunately, this proves not to be the case, for while information theory sets limits on how well one might do, it sets no limits on how poorly. The scaling of the GMUR as a function of the number of sources is a good example. The information needed to localize $N$ galactic sources is equal to

\begin{equation}
I_\Delta(N) = \log_2\left(\frac{V_{7N}}{\Delta V_{7N}}\right)
\end{equation}
where $\Delta V_{7N}$ is given by (\ref{vol}) and $V_{7N} = V_{7}^N$ is the volume of the $7N$ dimensional parameter space. Requiring that $I_\Delta(N) < I_S(N)$ yields a lower bound on the geometric mean of the parameter uncertainties (GMU) of

\begin{equation}\label{inf}
{\rm GMU} > \frac{V_7^7}{\left(1+P_s/P_n\right)^{BT/7N}} \, .
\end{equation}

Consider first the case of $N$ isolated binaries of similar brightness. The power ratio $P_s/P_n$ will be independent of $N$, and the bandwidth will be the sum of the bandwidths of each signal, so $B$ will grow linearly with $N$. Thus, equation (\ref{inf}) implies that for isolated binaries, the parameter uncertainties will be independent of the number of sources. This is indeed the case, as the Fisher information matrix is block diagonal for isolated sources. Now consider the case of $N$ overlapping binaries sharing the same fixed bandwidth $B$. The power $P_s$ will grow approximately linearly with $N$, leading to the prediction that the GMU must grow faster than $(1/N)^{(1/N)}$. The actual scaling seen in Figures~\ref{GMUR_vs_number_of_binaries_one_year_multiple_freqs} and \ref{GMUR_vs_number_of_binaries} tell a different story, with the GMUR increasing as $\sim N^N$. Thus, while information theory provides a lower bound on how quickly the parameter uncertainties must increase with the number of overlapping sources, the bound is too weak to be of any real use. The weakness of the bound is probably related to how poorly the information for multiple overlapping binaries is encoded.

\section{Discussion}\label{concl}

We have found that source confusion will significantly impact LISA's ability to resolve individual galactic binaries.  We found that source confusion affects parameter estimation fairly uniformly across the seven paramters that describe a galactic binary.

The two most significant findings were that source confusion grows exponentially with the number of correlated sources, and that source confusion decreases rapidly with time of observation.

Source confusion will be a significant problem in the frequency range between $0.01$ and $3$ mHz, where it is estimated that there are upwards of $10^8$ galactic binary systems. The decrease in the parameter uncertainties with time of observation is far greater than the usual $1/\sqrt{T_{obs}}$ decay that occurs when competing against stationary, Gaussian noise. The fast improvement is due to the $1/T_{obs}$ decrease in the source cross-correlation, and the sensitive dependence of the parameter uncertainties on the degree of cross-correlation.

Our findings suggest that earlier work that treated the galactic background as stationary, Gaussian noise should be revisited, and that every effort should be made to extend the LISA mission lifetime beyond its nominal 3 year duration.

\end{document}